\documentstyle[12pt, openbib]{article}

\title{ Two phase transitions in  (s+id)-wave
Bardeen-Cooper-Schrieffer  
superconductivity\thanks{Ref: J. of Phys.: Cond. Mat. 10 (1998) L319}}

\author{Angsula Ghosh and Sadhan K Adhikari\thanks{John Simon Guggenheim 
Memorial Foundation Fellow}\\
Instituto de F\'{\i}sica Te\'orica, Universidade Estadual Paulista,\\
01.405-900 S\~ao Paulo, S\~ao Paulo, Brazil\\}

\date{\today}
\begin{document}
\maketitle

\begin{abstract}We establish universal behavior in temperature 
dependencies of some observables in ${\it (s+id)}$-wave BCS
superconductivity in the presence of a weak $s$ wave. There also 
could appear  a  second second-order phase transition.  As temperature is
lowered past the usual critical temperature 
$T_c$, a less ordered superconducting phase is created  in $d$
wave, which changes to a more ordered  phase in $(s+id)$ wave at $T_{c1}$ 
($< T_c$).
The presence of two phase transitions manifest in two jumps in specific heat
at $T_c$ and $T_{c1}$.  The temperature  dependencies of susceptibility,
penetration depth, and thermal conductivity also confirm the new phase
transition.


\end{abstract} 


\newpage

The Bardeen-Cooper-Schrieffer (BCS) theory of superconductivity
\cite{e,t} has been successfully applied  to different systems in pure
angular momentum states such as $s$, $p$, and $d$ waves.  However, the
unconventional high-$T_c$ superconductors \cite{hi} with a high critical
temperature $T_c$ have  a complicated lattice structure with extended and/or
mixed symmetry for the order parameter \cite{n}.  Some of the high-$T_c$
materials have singlet $d$-wave Cooper pairs  and the order parameter has
$d_{x^2-y^2}$ symmetry in two dimensions \cite{n}.  Recent measurements
\cite{h} of the penetration depth $\lambda(T)$   and superconducting specific
heat at different temperatures $T$  and related theoretical analysis
\cite{t1,c} also support this point of view.  In some cases there is  the
signature  of an extended $s$- or $d$-wave symmetry.  
The possibility of a mixed $(s-d)$-wave symmetry was suggested
sometime ago by Ruckenstein et al. and Kotliar \cite{6}. There are
experimental evidences of mixed $s$- and $d$-wave symmetry in compounds such
as YBa$_2$Cu$_3$O$_7$ (YBCO) \cite{5}, and Bi$_2$Sr$_2$CaCu$_2$O$_{8+x}$
\cite{7}, where an [$s+\exp(i\theta)d$] symmetry is applicable.
  Recently, this idea
has been explored to explain the NMR data in the superconductor YBCO and the
Josephson critical current observed in YBCO-SNS and YBCO-Pb junctions
\cite{8}.  There have also been certain recent theoretical studies using
mixed $s$- and $d$-wave symmetries \cite{9} and  it was noted that it is more
likely to realize a stable mixed $s+id$ state than a $s+d$ state considering
different couplings and lattice symmetries.

It is quite natural that the Cooper electrons  might interact in both $s$ and
$d$ waves with different couplings. In the presence of simple central
potentials the Cooper problem separates in its decoupled $s$- and $d$-wave
components. The same decoupling occurs in a linear Schr\"odinger equation.
However, in the  nonlinear BCS theory, the presence of both $s$- and $d$-wave
components in the interaction would lead to an order parameter of mixed
symmetry and consequently a coupled set of BCS equations. The symmetry of the
order parameter is to be specified in order to solve this coupled set of
equations.

The normal state of  most high-$T_c$ materials  has not been satisfactorily
understood and there are controversies about the appropriate microscopic
hamiltonian and  pairing mechanism \cite{n,c}.  Despite this, we study the
$(s+id)$-symmetry case of the order parameter using the weak-coupling
microscopic BCS theory based on the Fermi liquid model to extract some
model-independent properties of such a description.  For a weaker $s$-wave
admixture, quite unexpectedly, we find another second-order phase transition
at $T=T_{c1}<T_c$, where the superconducting phase changes from a pure
$d$-wave state for $T>T_{c1}$ to a mixed $(s+id)$-wave  state for $T<T_{c1}$.
The specific heat exhibits two jumps at the transition points $T=T_{c1}$ and
$T=T_c$.  The temperature dependencies of the superconducting specific heat,
susceptibility, penetration depth and thermal conductivity change drastically
at $T=T_{c1}$ from power-law behavior (typical to $d$ state with node(s) in
the order parameter on the Fermi surface)  for $T>T_{c1}$ to exponential
behavior (typical to $s$ state with no nodes) for $T<T_{c1}$. The order
parameter for the present ($s+id$) wave  does not have a node on the Fermi
surface for $T<T_{c1}$ and it behaves like a modified/extended $s$-wave one.
The observables for the normal state are closer to the superconducting $l =2
$ state than to those for the superconducting $l=0$ state \cite{c}.
Consequently, superconductivity in $s$ wave is more pronounced than in $d$
wave.  Hence as temperature decreases the system passes from the normal state
to a ``less" superconducting $d$-wave state  at $T=T_c$ and then to a ``more"
superconducting extended  $s$-wave state at $T=T_{c1}$ signaling a second
phase transition.

We consider a system of $N$ superconducting electrons  under the action of a
purely attractive   two-electron potential in partial wave $l$ (=0,2):
\begin{equation}
V_{ \bf pq}=-  \sum_{l=0,2}V_l
\cos (l\theta_p)\cos(l\theta_q)  \label{pot}
\end{equation}   
where $\theta_p$ is the angle of momentum vector ${\bf p}$.

Potential (\ref{pot}) for a arbitrary small $V_l$  leads to Cooper pairing
instability at zero temperature in  even (odd) angular momentum states for
spin-singlet (triplet) state.  The Cooper-pair problem for two electrons
above the filled Fermi sea is given by
\cite{c}
\begin{equation} 
V_l^{-1}= \sum _{{\bf q}(q>1)}   \cos ^2(l\theta) (2\epsilon_q-\hat
E_l)^{-1}
\label{2}
\end{equation}
with the Cooper binding $C_l=2-\hat E_l$.  
Here   $\epsilon_q =\hbar^2 q^2/2m$ with $m$  the mass of an electron and
the ${\bf q}$-summation  is evaluated according to
\begin{equation} \sum _{\bf q} \to \frac {N} {4\pi} 
\int  d\epsilon_ q  d\theta 
\equiv \frac 
{N}{4\pi}
\int_0^\infty 
d\epsilon_ q \int_0^{2\pi}d\theta.\label{4}\end{equation}
Unless the units of the variables
are explicitly mentioned, in this work all energy   variables are expressed
in units of $E_F$, such that  $T \equiv T/T_F$,  $E_{\bf q}\equiv E_{\bf
q}/E_F$, $E_F=k_B=1,$ etc, where $T_F$ ($E_F$) is Fermi temperature (energy)
and $k_B$ the Boltzmann constant.

We consider a weak-coupling renormalized BCS model  in two dimensions with
$(s+id)$ symmetry.  At a finite $T$, one has  the following BCS  equation
\begin{eqnarray}
\Delta_{\bf p}& =& -\sum_{\bf q} V_{\bf pq}\frac{\Delta_{ \bf q}}{2E_{\bf
q}}\tanh
\frac{E_{\bf q} }{2T}  \label{130} \end{eqnarray}
 with $E_{\bf q} = [(\epsilon_q - \mu )^2 + |\Delta_{\bf q}|^{2}]^ {1/2}.$
The order parameter $\Delta _{\bf q}$ has the following anisotropic form:
$\Delta _{\bf q} \equiv \Delta_0+i \Delta_2 \cos(2 \theta),$ where
$\Delta_l$'s  are dimensionless.  The  BCS gap is defined by
$\Delta(T)=(\Delta_ 0^2+\Delta_2^2/2)^{1/2}$, which is the root-mean-square
average of $\Delta _{\bf q}$ on the Fermi surface.  Using the above form of
$\Delta_{\bf q}$ and potential (\ref{pot}), (\ref{130}) becomes the
following coupled set of BCS equations for $l=0$ and 2
\begin{equation}
\frac{1}{V_l}=\sum_{\bf q}\cos^2(l\theta)\frac{1}{2E_{\bf q}}\tanh
\frac{E_{\bf q}}{2T}\label{131}
\end{equation}
where the coupling is introduced through $E_{\bf q}$.

Using  (\ref{2}), set  (\ref{131}) of BCS equations  can  be explicitly
written in terms of  Cooper bindings as follows:
\begin{equation}
\int    d\theta 
\cos^2(l\theta) \biggr[\int_1^\infty 
\frac{2d\epsilon_q}{2\epsilon_q-\hat E_l}-
\int _0^\infty
\frac{d\epsilon_q}{E_{\bf q}}\tanh \frac{E_{\bf q}}{2T} \biggr]=0.\label{15a}
\end{equation}
The two terms in the BCS equation  (\ref{15a}) have ultraviolet divergence.
However, the difference between these two terms is finite.  BCS model
(\ref{15a}) is  independent of 
coupling $V_l$, is governed by Cooper binding
$C_{l}$, and has some advantages \cite{c}.  Firstly, no energy cut-off is
needed in this equation.  This is why model  (\ref{15a}) is  called
renormalized \cite{c,re}.  Secondly, this model leads to an increased $T_c$
in the weak-coupling limit, appropriate for some high-$T_c$ materials
\cite{c}.  Here,  we use the renormalized model for convenience. Otherwise,
it has no effect on our conclusions and the same analysis can be performed in
the standard  BCS model with cut off.

The  specific heat per particle  is given by \cite{t}
\begin{equation}
C(T)= \frac{2}{NT^2}\sum_{\bf q}   f_{\bf q}(1-f_{\bf q})
\left( E_{\bf q}^2-\frac{1}{2}T\frac{d|\Delta_{\bf q}|^2}{dT} \right)
\label{sp} 
\end{equation} 
where $f_{\bf q}=1/(1+\exp( E_{\bf q}/ T))$.  The spin-susceptibility $\chi$
is defined by
\cite{c}
\begin{equation}
\chi(T)= \frac{2\mu_N^2}{T}\sum_{\bf q}f_{\bf q}(1-f_{\bf q})
\end{equation}
where $\mu_N$ is the  nuclear magneton.  The penetration depth $\lambda$ is
defined by \cite{t}
\begin{equation}
\lambda^{-2}(T) = \lambda^{-2}(0)\left[1- \frac{2}{NT}
\sum_{\bf q} f_{\bf q}(1-f_{\bf q})\right].
\end{equation}  
The superconducting to normal thermal conductivity ratio $K_s(T)/K_n(T)$ is
defined by \cite{c}
\begin{equation}
\frac {K_s(T)}{K_n(T)} = \frac {\sum_{\bf q} (\epsilon_q -1)f_{\bf q}(1-f_
{\bf q})E_{\bf q} }{\sum_{\bf q} (\epsilon_q -1)^2 f_{\bf q}(1-f_{\bf q})}.
\end{equation}

We solved  the coupled set of equations (\ref{15a}) numerically and
calculated the  gaps $\Delta_0$ and $\Delta_2$ at various temperatures for
$T<T_c$.  The corresponding BCS gap $\Delta(T)$ was also calculated. For a 
very weak $s$-wave ($d$-wave) interaction the only possible solution 
corresponds to $\Delta_0 =0$ ($\Delta_2 =0$). In order to have a coupling
between $s$ and $d$ waves both the interaction potentials  are to be
reasonable. We
have 
studied the solution only when a coupling between the two equations is
allowed. In this domain we 
have kept the $d$-wave coupling stronger than $s$-wave coupling, so that as
temperature is lowered past $T_c$ a superconducting phase in $d$ wave 
appears.  In Fig.
1 we plot the temperature dependencies of different $\Delta$'s for the
following two sets of $s$-$d$ mixing corresponding to (1) $C_{0}= 0.0006$,
$C_{2}=0.001$, (full line) and (2) $C_{0}= 0.00085$, $C_{2}= 0.001$ (dashed
line),  referred to as models sd1 and sd2, respectively.  For a
superconductor with $T_F=5000$ K, the largest of these Cooper bindings
$C_{2}$ is 5 K. The smallness of this binding guarantees the weak-coupling
limit, where the BCS model should provide a good description.  In both cases
the parameter $\Delta_2$ is suppressed in the presence of a non-zero
$\Delta_0$. However, the BCS gap $\Delta(T)$ has the same form as in the case
of pure $s$ and $d$ waves.  In model sd1 (sd2)  $\Delta(0)/T_c$ = 1.535
(1.644), $T_c =$ 0.0266     (0.0266), $T_{c1} = 0.01065$     (0.0206).  For 
a pure $s$  ($d$) wave $\Delta(0)/T_c$ = 1.764 (1.513) \cite{c}. At
$T=0$ the order parameter has  $s$- and $d$-wave components and we find as
$T$ increases both components  decrease  and for $T\ge T_{c1} $ the $s$-wave
component vanishes and one is left with a pure $d$-wave component, which
vanishes at $T= T_c$.

In order to substantiate the claim of the second phase transition at
$T=T_{c1}$, we study the temperature dependence of specific heat in some
detail. The different specific heats are plotted in Fig. 2.  With this
two-step transition, the superconducting specific heat exhibits  a very
unexpected peculiar behavior.  In both models the specific heat exhibits two
jumps $-$ one at $T_c$ and another at $T_{c1}$. From  (\ref{sp}) and Fig.
1 we see that the temperature derivative of $|\Delta_{\bf q}|^2$ has
discontinuities at $T_c$ and $T_{c1}$ due to the vanishing of $\Delta_2$ and
$\Delta_0$, respectively, responsible for the two jumps in specific heat.
For $T_c > T > T_{c1}$, the specific heat exhibits typical $d$-wave power-law
behavior $C_s(T)/C_n(T_c)= 2(T/T_c)^2$ found in recent studies \cite{c}. For
$T<T_c$, we find an exponential behavior.   
Two jumps in specific heat have
been observed recently  in certain superconducting compounds
which suggest the existence of a coupled $s+id$ phase
\cite{exp}.

Next we study the temperature dependencies of spin susceptibility,
penetration depth, and thermal conductivity which we exhibit in Figs. 3 $-$ 5
where  we also plot the results for pure $s$ and $d$ waves from Ref. \cite{c}
for comparison.  In all cases  $d$-wave-type power-law behavior is obtained
for $T_c>T>T_{c1}$. We obtain  in $d$ wave $K_s(T)\approx K_n(T)(T/T_c)^
{1.2}$ and $\chi_s(T)/\chi_n(T_c) \approx (T/T_c)^{1.3}$ \cite{c}.
  For $T<T_{c1}$, there is no node in the present order
parameter on the Fermi surface and one has a typical extended $s$-state
behavior.  A passage from $d$  to extended $s$ state at $T_{c1}$ represents
an increase in order and hence an increase in superconductivity \cite{c}.  As
temperature decreases, the system passes from the normal state to a $d$-wave
state  at $T=T_c$ and then to an extended $s$-wave state at $T=T_{c1}$
signaling a second phase transition.

In conclusion, we have studied the $(s+id$)-wave superconductivity employing
a renormalized BCS model in two dimensions and confirmed a  second-order
phase transition at  $T=T_{c1}$ in the presence of a weaker $s$ wave.  We
have kept the $s$- and $d$-wave couplings in such a domain that a coupled 
$(s+id)$-wave solution is allowed. As
temperature is lowered past the  first critical temperature $T_c$, a weaker
(less ordered) superconducting phase is created  in $d$ wave, which changes
to a stronger (more ordered) superconducting phase in $(s+id)$ wave at
$T_{c1}$.     The
$(s+id)$-wave state is similar to an  extended $s$-wave state with no node in
the order parameter. The phase transition at $T_{c1}$ is also marked by
power-law (exponential) temperature dependencies of $ C(T), \chi(T)$, $\Delta
\lambda (T)$ and $K(T)$ for $T > T_{c1}$ ($<T_{c1} $).  A similar
second-order phase transition may occur for some other types of mixtures of
angular momentum states.

We thank Dr Haranath Ghosh for a very helpful discussion and Conselho
Nacional de Desenvolvimento Cient\'{\i}fico e Tecnol\'ogico and Funda\c c\~ao
de Amparo \`a Pesquisa do Estado de S\~ao Paulo for financial support.

{\bf Figure Captions:}
\vskip 1cm

1. The  $s$- and $d$-wave parameters
$\Delta_0$,  $\Delta_2$, and  BCS gap 
$\Delta(T)$ at different temperatures for $(s+id)$-wave models 
sd1 (full line) and sd2 (dashed line)
described in the text with different mixtures of $s$ and $d$ waves.

2. Specific heat ratio $C(T)/C_n(T_c)$ versus $T/T_c$  for models sd1 
(full line) and 
sd2 (dashed line). The dotted line represents the pure $d$-wave result 
 from Ref. \cite{c}   for comparison.

3. Spin susceptibility  ratio $\chi_s(T)/\chi(T_c)$ versus $T/T_c$  
for models sd1 
(full line) and 
sd2 (dashed line). The dotted lines represent the pure 
$s$ and $d$-wave results 
 from Ref. \cite{c}   for comparison.

 4. Penetration depth  ratio $\Delta \lambda(T)\equiv
 [\lambda(T)-\lambda(0)]/\lambda(0)$ versus $T/T_c$  
for models sd1 
(full line) and 
sd2 (dashed line). The dotted lines represent the pure 
$s$ and $d$-wave results 
 from Ref. \cite{c}   for comparison.

5. Thermal conductivity   ratio $K_s(T)/K_n(T)$ versus $T/T_c$  
for models sd1 
(full line) and 
sd2 (dashed line). The short dashed lines represent the pure 
$s$ and $d$-wave results 
 from Ref. \cite{c}   for comparison.

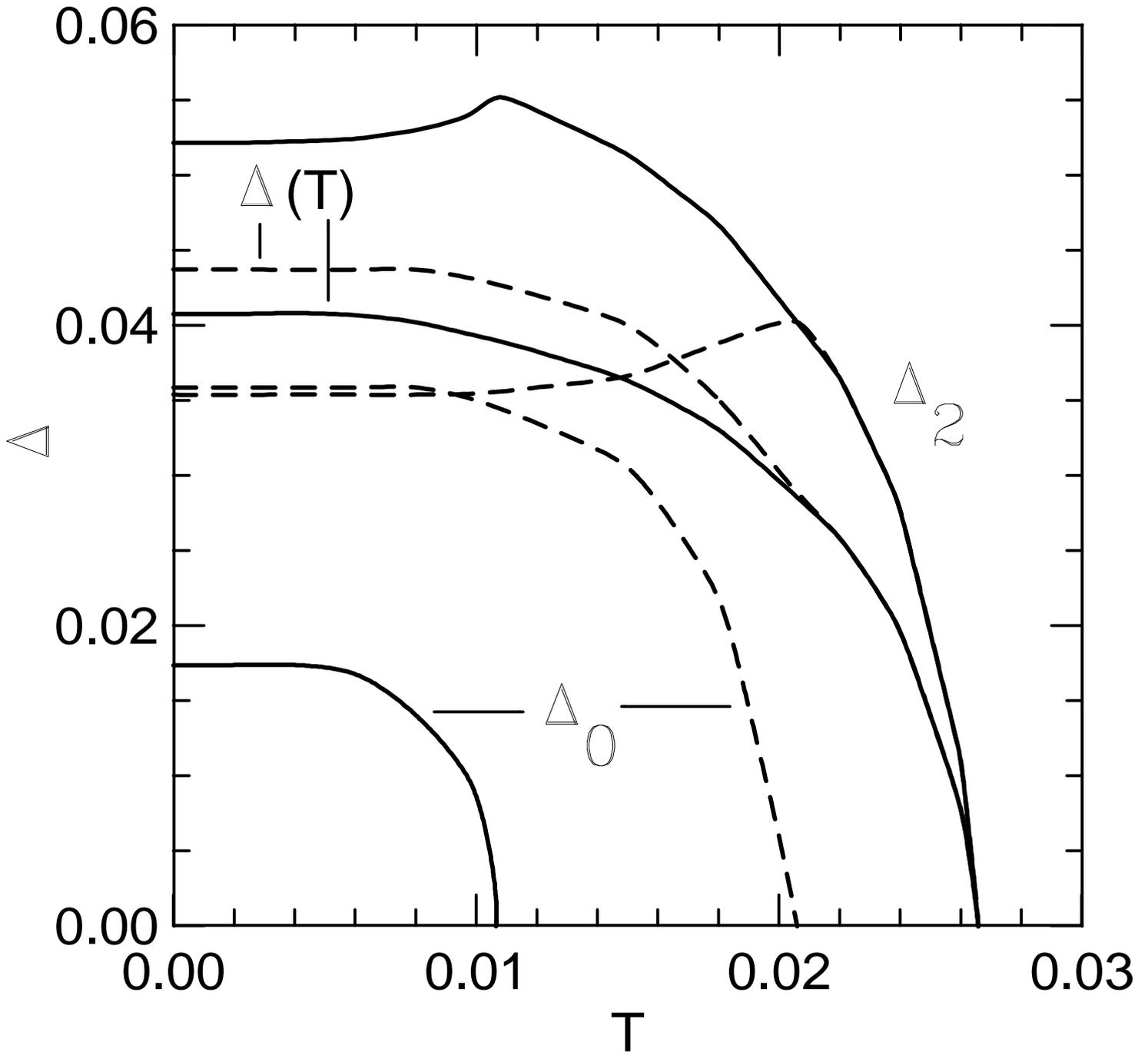
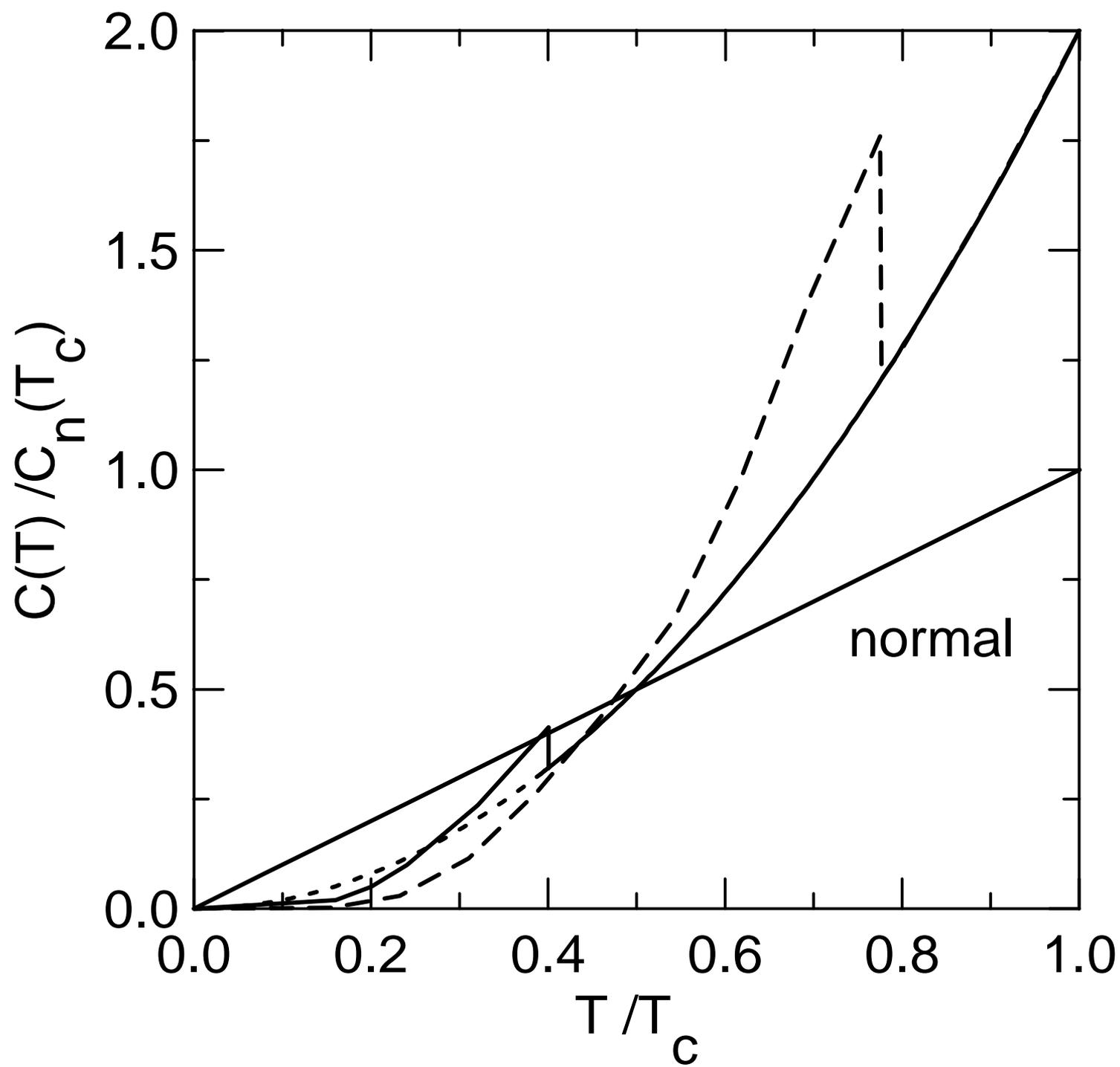
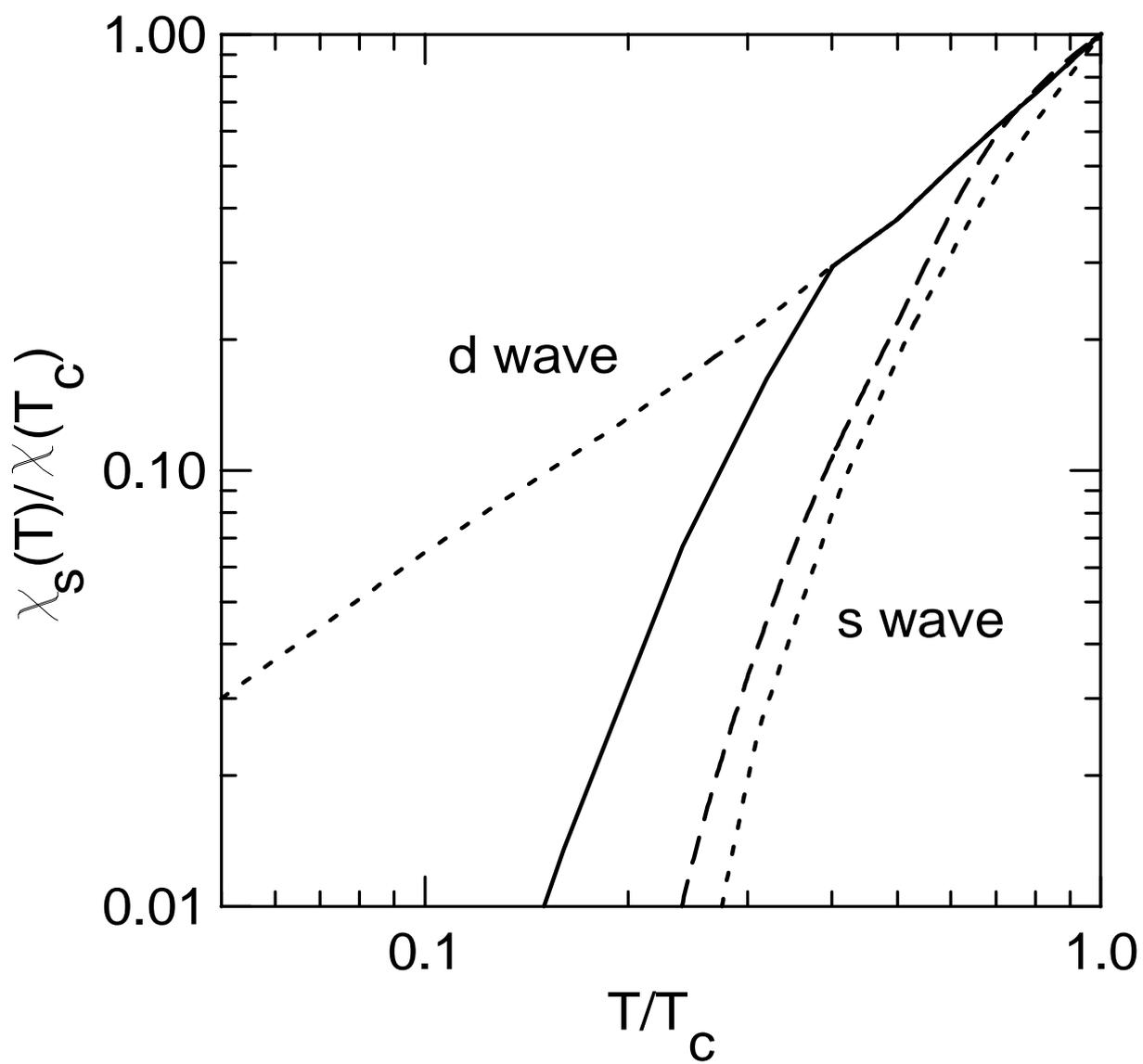
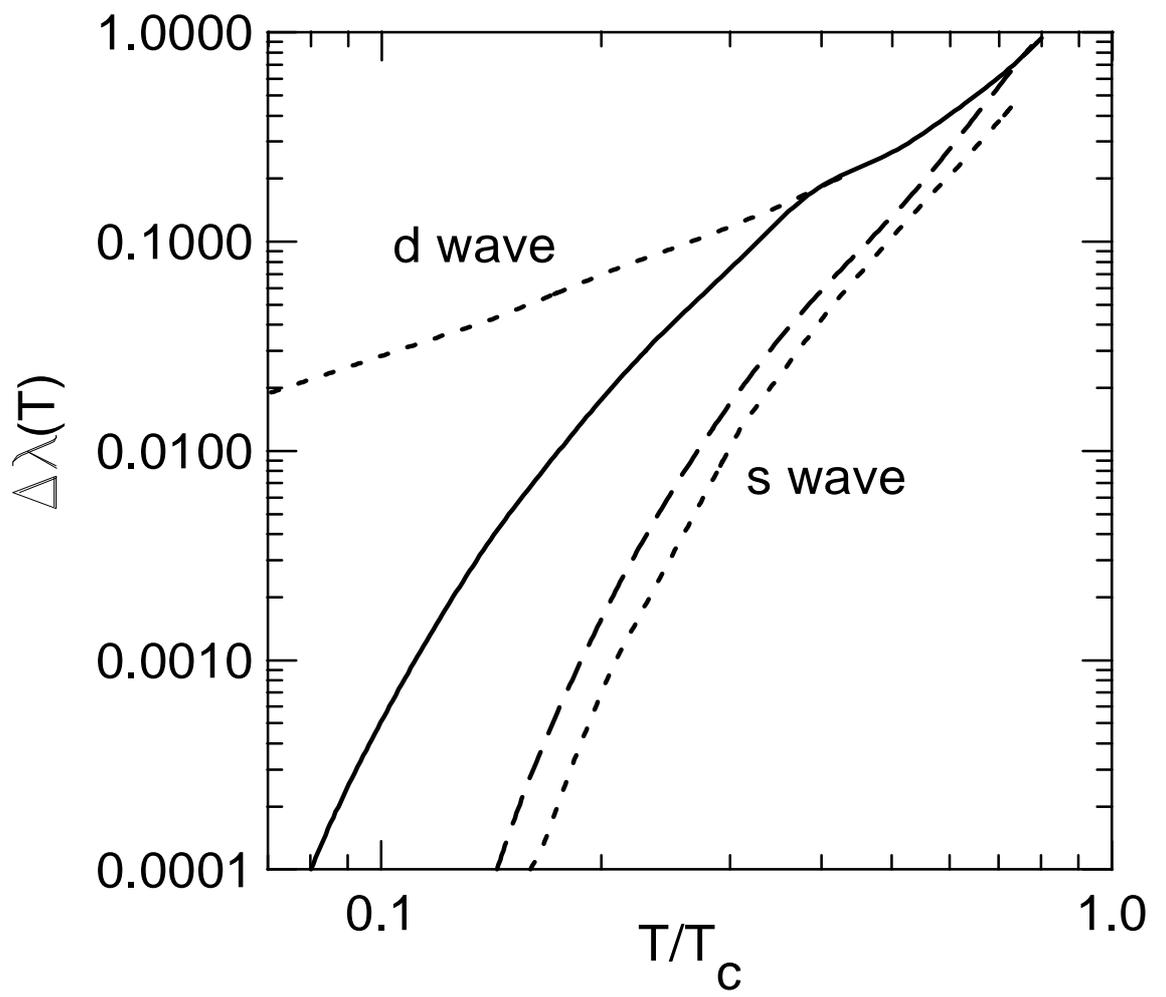
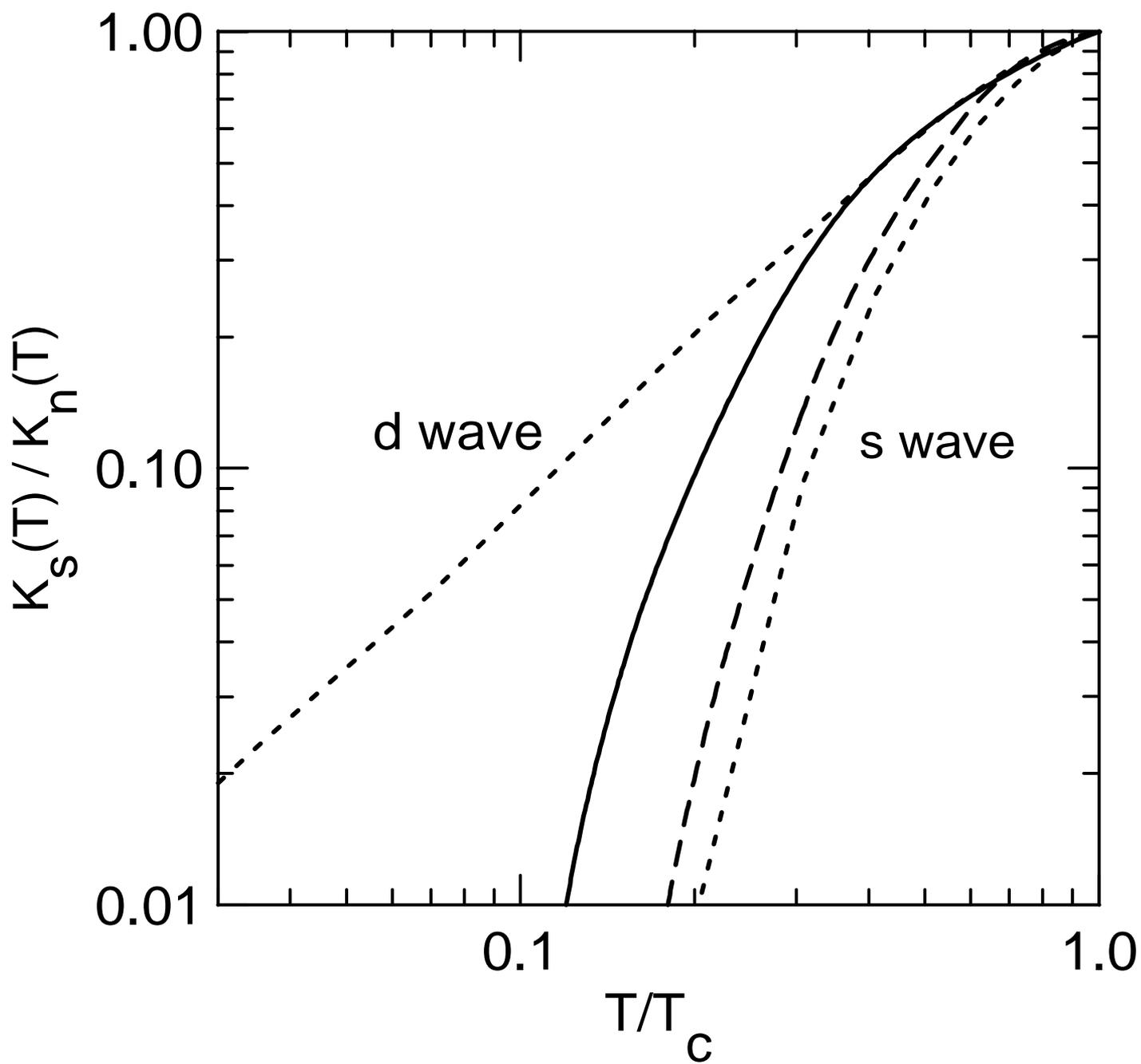
\end{document}